\documentclass[11pt,twoside]{article}
\usepackage{asp2014}

\aspSuppressVolSlug
\resetcounters

\begin{document}

\title{The International Virtual Observatory Alliance (IVOA)  in 2020}

\author{G.~Bruce~Berriman$^1$, on behalf of the IVOA Executive Committee, the IVOA Technical Coordination Group and the IVOA Community}
\affil{$^1$Caltech/IPAC-NExScI, Pasadena, CA, 91125; \email{gbb@ipac.caltech.edu}}


\paperauthor{G.~Bruce Berriman}{gbb@ipac.caltech.edu}{0000-0001-8388-534X}{Caltech}{IPAC/NExScI}{Pasadena}{CA}{91125}{USA}



  
\begin{abstract}

The International Virtual Observatory Alliance (IVOA) develops the technical standards needed for seamless discovery of and access to astronomy data worldwide, according to the Findable, Accessible, Interoperable and Reusable (FAIR) principles, with the goal of realizing the Virtual Observatory (VO). There are 21 member organizations. The Netherlands VO applied for membership in 2020. Astronomical communities from other nations have shown interest in joining the IVOA. This paper describes the activities of the IVOA in 2020 and summarizes the May and November 2020 virtual "interoperability meetings." The May meeting was the first to be held online and the first to have over 200 registrants.
  
\end{abstract}

\section{Introduction: What is the IVOA?}
The International Virtual Observatory Alliance\footnote{\url{http://www.ivoa.net}} (IVOA) delivers the technical standards needed for seamless discovery and access to astronomy data, according to the Findable, Accessible, Interoperable and Reusable (FAIR) principles \citep{doi:10.1038/sdata.2016.18}. These standards are realizing the Virtual Observatory (VO), which is best thought of as an ecosystem of inter-operable tools and services that deliver metadata and data from archives worldwide. 

IVOA operates as a framework for discussing and sharing ideas and technology, for engaging astronomy projects, missions and researchers, and for promoting and publicising the VO. The IVOA has 21 national member organizations. The Netherlands VO applied for membership in 2020. The technical work is carried out by 6 Working Groups and 8 Interest Groups, all coordinated by the Technical Coordination Group (TCG). A new interest group, Radio Astronomy, was founded in 2020. A Committee for Science Priorities (CSP) recommends science priorities, and overall direction is provided by the IVOA Executive Committee. The IVOA participates in international cross-discipline initiatives to support development of the social and technical infrastructure to enable open sharing and re-use of data (e.g., the Research Data Alliance\footnote{https://rd-alliance.org/}; and the the General Conference of Weights and Measures initiative to redefine the International System of Units. \footnote{https://www.bipm.org/en/measurement-units/}

\section{IVOA Meetings}
The bulk of IVOA work centers on open bi-annual interoperability meetings, one in the northern Spring and one in the northern Fall. Since 2014, the Fall meeting has been held consecutively with the ADASS meeting. The Spring 2020 meeting was to have been held in Sydney, Australia \footnote{https://wiki.ivoa.net/twiki/bin/view/IVOA/InterOpMay2019}, and the Fall 2020 meeting in Granada, Spain \footnote{https://wiki.ivoa.net/twiki/bin/view/IVOA/InterOpMay2019}. The COVID-19 pandemic did, however, mandate that these meetings were held virtually.

The virtual nature of the meetings required substantial changes in format. To allow participants across 19 time zones to take part, sessions were staggered throughout the 24-hour day; all sessions were recorded and served through the CANFAR VOSpace service \footnote{vos://cadc.nrc.ca~vault/pdowler/ivoa/virtual2020a}. The discussion was captured and archived in text files created with the Etherpad application. While the staggered sessions did impose limits on participation, the two meetings had for first time over 200 registrations.

\section{IVOA Achievements and Impact in 2020}

The primary goals of the IVOA are to develop and evolve standards that provide common interfaces to discover and access data sets; support and advocate for the incorporation of these standards in the query infrastructures of archives and data centers; encourage the development of data analysis tools that use these standards; support professional outreach efforts; and support education and outreach efforts. The work is currently guided by two science goals recommended by the CSP:  support the emergent field of time domain astronomy (TDA) and multi-messenger astronomy (MMA), and support for multi-dimensional data, including  spectral and  temporal data cubes. The goals are driven by, on the one hand, the construction of dedicated time domain telescopes, and on the other hand, by the accelerating growth of large and complex data sets.

Two decades of continual effort has led to take-up of VO protocols in astronomy, along with a consequent maturity of the VO ecosystem. What follows highlights and illustrates the scale and scope of the IVOA's work in 2020; see the online proceedings for a complete account. 

\subsection{IVOA Protocols in Archives and Data Centers}

In large measure because of active engagement with data providers, IVOA protocols are now embedded in the architectures of all major archives and data centers. Here, we list just some examples from 2020. The ESASky client has integrated the Table Access Protocol (TAP) services from the European Southern Observatory (ESO), Canadian Astronomy Data Centre and the Mikulski Archive for Space Telescopes, as well as the Chandra source catalog v2. The European Space Agency (ESA) has also recently released a new VO-compliant ISO archive, while the Euclid archive is integrating TAP, DataLink and Simple Image Access Protocol v2 (SIAPv2). The ESA archive now offers access to all Gaia data releases through a TAP service. The NASA Exoplanet Science Institute released a Python-based TAP server that underpins queries to the NASA Exoplanet Archive, Keck Observatory Archive, and the future archive for the NEID extreme precision radial velocity spectrograph. In Australia, a national effort is integrating VO services into data centers:  the SkyMapper Southern Survey has delivered its Data Release 2, and CASDA is now serving data from four different surveys (RACS, EMU, WALLABY, GASKAP). The VO-compliant archive for Gran Telescopio Canarias has been released by the Spanish Virtual Observatory (SVO). The information management system for the Chinese Five-hundred-meter Aperture Spherical Radio Telescope single radio dish is built on IVOA protocols.  
  
\subsection{VO-compliant Data Access, Visualization and Analysis Tools}
New and upgraded VO clients and tools have been released in 2020.
Aladin Desktop version 11 emphasizes support for time domain astronomy. Clusterix 2.0, from the SVO, determines membership probabilities in stellar clusters based on proper-motion data; access to Gaia data is included.  The ESASky tool is now available in Chinese.  The IPAC Firefly tool data integration tool now support VOTAble, TAP, Hierarchical Progressive Surveys (HiPS), Multi-order Coverage (MOC) maps, and Datalink.  

\subsection{Time Domain and Multi-Messenger Astronomy} 

Driven by multi-messenger and time-domain astronomy, the IVOA is developing two standards aimed at coordinating rapid follow-up on ground-based telescopes. These protocols, the Object Visibility Simple Access Protocol (ObjVisSAP) and the Observation Locator Tabular Access Protocol (ObsLocTAP), have been endorsed by a Kavli-IAU international workshop on coordinating MMA follow-up \citep{2020arXiv200705546C}. There are plans to seek IAU endorsements of these standards in 2021.

Further, Version 2.0 of the Multi-Order Coverage (MOC) protocol, scheduled for completion in 2021, will provide generalization to support spatial and temporal sky coverage. One of its use cases is localization of gravitational wave events.

\subsection{Professional Outreach}

IVOA members organize and take part in tutorials and workshops aimed at demonstrating the use of IVOA protocols, to encourage their take-up, and to secure technical feedback on them. 

The NASA Virtual Observatory (NAVO) has held tutorials at several national AAS meetings on the use of Python to discover, access and process archival data. It plans to continue these workshops in 2021. 

ESAC/ESA held a Demonstrator workshop in September attended by representatives of 40 observatories to discuss VO protocols, and especially ObjVisSAP and ObsLocTAP. 

ESCAPE held a Provenance Workshop in September. It demonstrated the value of IVOA standards in creating provenance metadata for large and complex data sets.

\subsection{Education and Public Outreach}

The World Wide Telescope (WWT) is now managed by the American Astronomical Society (AAS), which is investing in making the architecture more sustainable. Part of this effort is to make HiPS maps accessible through the WWT; much of this development has been performed by China-VO.

The IVOA is preparing to contribute to the project "Astronomy From Archival data," (Dr Priya Prisan, PI) which responded to a call from the IAU Office of Astronomy for Development for educational and outreach proposals that seek to provide services that mitigate the effects of COVID-19.

\section{Prospects for 2021 and Beyond}
Astronomy is poised to enter the era of petabyte-scale data sets, with new telescopes such as the Rubin Observatory and the Square Kilometer Array expected to begin operations in the 2020s. New missions such as Euclid, Roman and SPHEREx will offer extraordinary new views on the Universe. The challenge for the IVOA is to build on the progress to date to develop standards that will support these new missions. These challenges will involve work on several fronts:

\begin{itemize}
    \item The IVOA has begun support for "science platforms," where analysis takes place close to remote, massive data sets, likely hosted on the cloud. The ESCAPE project has begun to integrate the VO into the European Open Science Cloud. NAVO has similarly begun efforts to develop science platforms for analyzing NASA data; VO protocols will be embedded in their architecture. A particular challenge for all science platforms will be to incorporate theory simulations, whose metadata are very complex. 
    \item The IVOA will support new data-type adoption, driven by the growth in size and complexity of data sets. For example, columnar storage formats for large data sets, such as Apache Parquet was a topic at the November meeting. The data products of instruments such as SPHEREx are driving upgrades to the Simple Spectral Access Protocol. Finally, there is a need to incorporate radio visibility data into VO standards.
    \item Python APIs will play an important role in science platforms. The IVOA is coordinating with Astropy to provide Python reference implementations via the PyVO client.
    \item The priorities described above will be supported by workshops and tutorials, e.g., Euro-VO will host a VO school and a Data Provider Forum in 2021.

\end{itemize}



\end{document}